\documentclass[prl,twocolumn,superscriptaddress]{revtex4}
\usepackage{amsmath}
\usepackage{graphicx}
\usepackage{amssymb}
\usepackage{amsthm, amscd}
\usepackage{amsfonts}
\usepackage{cancel}
\usepackage{times}
\usepackage{pstricks}
\usepackage{wasysym}
\usepackage{ulem}
\usepackage[dvips]{epsfig}
\usepackage{subfigure}
\usepackage{feynmf}

\usepackage{amssymb}

\newcommand{\dslash}{\not{\hbox{\kern-2pt $\partial$}}}
 
\newcommand{\bq}{\begin{equation}} 
\newcommand{\eq}{\end{equation}}
\newcommand{\bqa}{\begin{eqnarray}} 
\newcommand{\eqa}{\end{eqnarray}}
\newcommand{\nn}{\nonumber \\}
\newcommand{\bw}{\begin{widetext}}
\newcommand{\ew}{\end{widetext}}

\newcommand{\cmu}{ { \{ \mu \}  } }

\begin{document}


\title{
Quantum Renormalization Group and Holography
}

\author{Sung-Sik Lee$^{1,2}$\\
{\normalsize{$^1$Department of Physics $\&$ Astronomy, McMaster University,}}
{\normalsize{1280 Main St. W., Hamilton ON L8S 4M1, Canada}}\\
{\normalsize{$^2$Perimeter Institute for Theoretical Physics,}}
{\normalsize{31 Caroline St. N., Waterloo ON N2L 2Y5, Canada}}
}

\date{\today}

\begin{abstract}

{\it Quantum renormalization group} scheme provides
a microscopic understanding of holography 
through a general mapping 
between the beta functions of underlying quantum field theories
and the holographic actions in the bulk.
We show that 
the Einstein gravity
emerges as a long wavelength holographic description
for a matrix field theory which has no
other operator with finite scaling dimension
except for the energy-momentum tensor.
We also point out that 
holographic actions for general large $N$ matrix field theories 
respect the inversion symmetry 
along the radial direction in the bulk if 
the beta functions of single-trace operators 
are gradient flows with respect to the target space metric 
set by the beta functions of double-trace operators.

\end{abstract}

\maketitle

Renormalization group (RG) flow describes how
short distance fluctuations modify
coupling constants (coupling functions in general) 
as a system is probed at progressively larger length scales.
Although RG provides a general framework for quantum field theories\cite{WILSON,POLCHINSKI84,WETT}, 
it is of limited practical use for strongly coupled theories
due to the fact that one has to keep track of a large (often infinite) set of 
operators.

AdS/CFT correspondence\cite{MALDACENA} 
provides an alternative way of organizing RG 
which is tractable 
for a certain set of strongly coupled quantum field theories.
According to the dictionary of the conjecture\cite{GUBSER,WITTEN},
$D$-dimensional coupling functions of quantum field theories
become dynamical variables in a $(D+1)$-dimensional bulk space.
The radial direction in the bulk
plays the role of the length scale in RG. 
The saddle point solution of an action in the bulk
describes the evolution of the coupling functions 
along the radial direction,
which can be interpreted as RG flow\cite{EMIL,VERLINDE,HOLORG}.
Despite this natural interpretation, 
the connection between holography and RG
has been incomplete because the bulk fields are 
in general {\it dynamical and quantum} variables.
They are dynamical in the sense that the bulk action includes 
two-derivative terms along the radial direction, 
and quantum because bulk fields have non-trivial quantum fluctuations.
In conventional RG, on the contrary, 
coupling functions are 
non-dynamical and classical in the sense that
they obey first-order beta functions,
and an initial condition completely fixes 
the coupling functions at lower energy scales 
without any uncertainty. 

A precise connection between holography
and RG can be made via 
{\it quantum renormalization group}\cite{SLEE10,SLEE112}.
Unlike the conventional RG scheme, 
only a subset of operators is kept in quantum RG.
The price one has to pay is to promote  
the coupling functions to dynamical fields.
The partition function is given by
a sum over all possible RG paths 
for the coupling functions 
of the operators in the subset. 
The weight for each path is determined 
by an action for the scale-dependent dynamical sources.
In the context of matrix field theories, 
one needs to include only single-trace operators,
although multi-trace operators are generated
in the Wilsonian effective action\cite{BECCHI}.
In quantum RG, 
double-trace operators generated at each step 
of coarse graining become kinetic terms for the 
sources of single-trace operators, allowing
them to have non-trivial quantum fluctuations.
The role of double-trace operators in holography
was also emphasized in Refs. \cite{HEEMSKERK10,Faulkner1010}. 
Quantum RG allows one to establish precise connections 
between the beta functions of quantum field theories
and the bulk actions.
In this paper, we show that quantum gravity
can be derived from a matrix field theory via quantum RG.


Let us consider a large $N$ matrix field theory in the 't Hooft limit.
We consider a set of  primary single-trace operators $\{ O_n \}$ 
constructed from a trace of products of microscopic matrix fields, $\Phi_a$.
Any operator can be written as 
derivatives and multiplications of them.
In conventional RG, one has to include not only the
single-trace operators but also all multi-trace operators.

\begin{figure}[h!]
\centering
      \includegraphics[height=6cm,width=8cm]{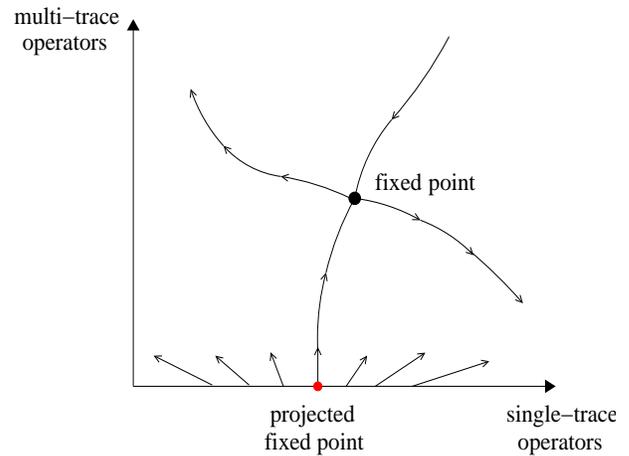}
\caption{
At fixed points, RG flow vanishes.
At projected fixed points, 
only the projected RG flow along the subspace of 
single-trace operators 
vanishes.
}
\label{fig:1}
\end{figure}

Quantum RG is formulated only in terms of the primary single-trace operators. 
The first step is to project fixed points to the subspace of the single-trace operators.
In general, there exists no real fixed point on the subspace because multi-trace operators are generated out of single-trace operators under RG flow. 
However, one can have a \textit{projected fixed point}.
It  refers to a theory at which RG flow is orthogonal to the subspace. 
In other words, a projected fixed point is a fixed point under the RG flow projected onto the subspace of single-trace operators. This is illustrated in Fig. \ref{fig:1}.

Let us consider a projected fixed point 
whose action $S_0[\Phi]$ is made of 
single-trace operators
in the $D$-dimensional Minkowski space.
%
To the theory, single-trace and multi-trace deformations can be added.
The generating function is given by
$Z  =  \int D \Phi ~ e^{ i ( S_0 + S_1 + S_2 )}$.
Here the single-trace deformation is written as
$S_1[ O_m; J^m]  =  N^2 \sum_m \int d^Dx J^m(x)  O_m$
with spacetime dependent sources $J^m(x)$.
$S_2[O_m,{\cal J}]$ is the multi-trace deformation 
which includes quadratic or higher order terms in $O_m$,
where ${\cal J}$ represents sources for the multi-trace operators.
We normalize the single-trace operators to be $O(1)$.

The multi-trace deformations can be removed by promoting the sources for the single-trace operators
to dynamical fields.
For this, we rewrite the generating function as
\bqa
Z[ J, {\cal J} ] & = & \int D \Phi ~~ 
\exp\left( i S_0 +i S_2\left[ -\frac{i}{N^2} \frac{\delta}{\delta J^m};  {\cal J}   \right] \right) 
\nn && \times
\exp ( i S_1[O_m, J^m] ),
\label{Z0}
\eqa
where every $O_m(x)$ in $S_2$ is replaced by
the functional derivative, 
$-\frac{i}{N^2} \frac{\delta}{\delta J^m(x)}$.
We introduce a pair of auxiliary fields
$j^{(0)m}$, $p_m^{(0)}$ for each single-trace  operator\cite{SLEE10,SLEE112}  to write
\bqa
Z[ J,{\cal J} ] & = & \int Dj^{(0)m} Dp_m^{(0)} D \Phi ~~ 
e^{ i S_0 + i N^2 \int d^Dx ~ p_m^{(0)} ( j^{(0)m} -  J^m ) }
\nn && \times
e^{ i S_2 \left[ -i/N^2 \delta/ \delta j^{(0)m} ;  {\cal J}  \right]} 
e^{i S_1[O_m; j^{(0)m}]  }. 
\label{Z01}
\eqa
Integrating  $j^{(0)m}$ by part,
the multi-trace terms are turned 
into a polynomial of $p_m^{(0)}$,
\bqa
Z[ J, {\cal J} ] & = & \int Dj^{(0)m} Dp_m^{(0)}  ~~ 
e^{ i N^2 \int d^Dx ~ p_m^{(0)} ( j^{(0)m} -  J^m ) }
\nn && \times
e^{i  S_2 \left[  -p_m^{(0)} ;  {\cal J}  \right] }
Z[ j^{(0)} ],
\label{Z02}
\eqa
where $Z[ j^{(0)} ]$ is the generating function 
with the single-trace deformation, $S_1[O_m; j^{(0)m}]$.
The original theory with multi-trace deformations
is mapped into a theory with only single-trace operators whose sources are dynamical.

Now we focus on $Z[ j^{(0)} ]$ and apply renormalization group procedure.
Under a coarse graining, high-energy modes are integrated out, 
and the UV-cut off is lowered by a factor of $e^{-dz}$. 
This renormalizes the deformation into 
$S_1[O_m; j^{(0)m}] + \delta S$, where
\bqa
&& \delta S[ O_n; j^{(0)n}]  =  dz N^2 \int d^Dx \Bigg\{  
{\cal L}_c(x;j^{(0)}] \nn
&& - \beta^{m}(x;j^{(0)}] O_m  
+ \frac{ G^{m n \cmu}(x;j^{(0)}] }{2} 
 O_m 
 \partial_\cmu O_n  
 \Bigg\}.
\label{dS}
\eqa
Here $f(x;j^{(0)}]$ denotes a function that
depends on $j^{(0)}(x)$ 
and their derivatives at position $x$.
${\cal L}_c(x;j^{(0)}]$ is the Casimir energy that is generated out of integrating out 
high-energy modes.
It can be viewed as the quantum correction to the identity operator. 
$\beta^{m}(x;j^{(0)}]$ 
represents the beta functional for the single-trace operators
\footnote{Descendants of $O_m$ can be removed by integration by part.}.
Because $S_0$ is a projected fixed point, 
$\beta^{m}$ (but not $G^{m n \cmu}$) vanishes at $j^{(0)}=0$, 
and can be expanded as
$\beta^m(x;j^{(0)}]  =  \Delta^m_n j^{(0)n} + O( (j^{(0)})^2 )$.
If there exists a single-trace operator with a scaling dimension $\Delta_O$ 
under the projected RG flow,
$\Delta^m_n$ has an eigenvalue $\Delta_O-D$.
$G^{m n \cmu}(x;j^{(0)}]$ is the source for double-trace operators
generated from quantum corrections.
$\cmu$ denotes a series of derivative, $(\partial_{\mu_1} \partial_{\mu_2} ...)$.
No higher-trace operators are generated to the order of $dz$.
Following the same steps as in Eqs. (\ref{Z0}) - (\ref{Z02}),
another set of auxiliary fields are introduced 
to remove the double-trace operators as
\bqa
Z[ j^{(0)}] & = & \int Dj^{(1)n} Dp_n^{(1)}  ~~ 
e^{ i  N^2 \int d^Dx ~ p_n^{(1)} ( j^{(1)n} -  j^{(0)n} ) }
\nn && \times
e^{ i  \delta S[ -p^{(1)}_n; j^{(0)n}]}
Z[ j^{(1)} ] .
\label{Z3}
\eqa
It can be explicitly checked that $S_1+\delta S$ 
is reproduced as the auxiliary fields are integrated out.

By iterating these steps, one can write the generating function 
as a functional integration of the auxiliary fields introduced at each step of coarse graining,
\bqa
Z[ j^{(0)} ] & = & \int \Pi_{l=1}^L Dj^{(l)n} Dp_n^{(l)} ~~ 
e^{ iS_B } Z[ j^{(L)} ], 
\label{Z4}
\eqa
where
\bqa
S_B & = &
N^2 dz \sum_{l=1}^L \int d^Dx  \Bigl\{ 
  p_n^{(l)} \frac{ j^{(l)n} - j^{(l-1)n} }{dz}  \nn
&& + \delta S[ -p_n^{(l)}; j^{(l-1)n} ] \Bigr\} 
\eqa
In the $dz \rightarrow 0$ limit, 
$j^{(l)n}(x)$, $p^{(l)}_n(x)$
become $(D+1)$-dimensional fields, $j^{n}(x,z)$, $p_n(x,z)$ 
with $z=l dz$.
The bulk action is written as
\bqa
S_B & = &
N^2 \int_0^{z^*} dz \int d^Dx ~  \Bigl\{ 
 p_n ( \partial_z j^{n} ) 
+  {\cal L}_c(x;j]  \nn
&&  + \beta^{m}(x;j] p_m 
+ \frac{ G^{m n \cmu}(x;j] }{2}  p_m \partial_\cmu  p_n  
\Bigr\}, 
\label{Sb}
\eqa
where $z^*=L dz$ is an IR scale at which we stop the RG procedure.
Without loss of generality, we can take $z^*=\infty$. 
If the scale $z$ is interpreted as a `time',
the dynamical source $j$ and the operator field $p$ become canonically conjugate to each other.
The Casimir energy, ${\cal L}_c$ becomes the potential `energy' of the source fields $j$,
and the quadratic term in $p$ becomes the kinetic `energy'.
The fact that sources become dynamical in the bulk is also natural from
the point of view of string theory in that
dynamical closed strings provide sources for open strings
which define field theory on D-branes\cite{ELIAS}.

Now we consider a scale-reversal (SR) transformation :
$j^n(x,z) \rightarrow j^n(x,z^*-z),
p_n(x,z) \rightarrow -p_n(x,z^*-z)$.
Since RG flow is irreversible,
one naively expects that the bulk action should always break the SR symmetry.
However, SR symmetric bulk actions can still describe irreversible 
RG flows because of  a boundary at the UV cut-off scale
(and also an IR boundary if there is an IR cut-off as well),
which explicitly break the SR symmetry.
In the bulk action, only the third term in Eq. (\ref{Sb}) breaks the symmetry.
It turns out that the SR-symmetry breaking term in the bulk
can be removed if there exists a $D$-dimensional functional $c[j(x)]$
which generates the projected RG flow of single-trace operators
as a gradient flow,
\bqa
\beta^{m } (x;j] = G^{m n \cmu}(x;j] \partial_\cmu 
\frac{ \delta c[j] }{ \delta j^{n}(x) },
\eqa
where $G^{m n \cmu}(x;j]$ plays the role of a `super-metric\rq{}
in the space of single-trace operators.
In this case, the conjugate momentum can be shifted as
$p^n \rightarrow p^n -  \frac{\delta c}{\delta j_n}$,
and the SR odd term becomes a boundary terms, $N^2 \left( c[j(x,0)] - c[j(x,z^*)] \right)$.
Then the bulk action is written as
$S_B  = 
N^2 \int dz   \Bigl\{ 
\int d^Dx ~ p_n ( \partial_z j^{n} ) - H \Bigr\}$,
where the `Hamiltonian' is given by
\bqa
H & = &   -\frac{1}{2}  
{\boldsymbol p}^T
\cdot
{\bf G} 
\cdot
{\boldsymbol p } 
+ \frac{1}{2} {\boldsymbol  \beta}^T 
\cdot
{\bf G}^{-1} 
\cdot
{\boldsymbol  \beta} 
-  \int d^Dx {\cal L}_c
\label{Sb1}
\eqa
which respects the SR symmetry.
Here ${\boldsymbol  p}$, ${\boldsymbol  \beta}$ are understood as vectors 
whose indices run over $m$ and $x$
(${\boldsymbol  p}^T$, ${\boldsymbol  \beta}^T$ are their transposes) 
and ${\bf G}$ is a matrix.
The second term on the r.h.s. of Eq. (\ref{Sb1}) along with the quadratic term 
of the Casimir energy determines the mass
of the source fields.
To the quadratic order in $j$, it becomes
$
\frac{1}{2} {\boldsymbol  \beta}^T 
\cdot
{\bf G}^{-1} 
\cdot
{\boldsymbol  \beta} 
= \frac{1}{2} {\bf j}^T
\cdot {\boldsymbol \Delta}^T 
\cdot
{\bf G}^{-1} 
\cdot {\boldsymbol \Delta} 
\cdot 
{\bf j}$.
This explains 
why the scaling dimension of an operator
determines the mass of the corresponding field in the bulk. 
Note that the mass is positive when ${\bf G}$ is positive.
In this case, the kinetic term has the `wrong' sign.
This implies that the radial direction is space-like not time-like.


As a concrete example, 
let we consider a projected fixed point
of a matrix field theory 
where the single-trace energy-momentum 
tensor is the only operator that has finite
scaling dimension under projected RG flow.
All other single-trace operators have infinite scaling dimensions 
and they instantly die out if they are generated under RG flow.
Therefore a general single-trace action 
is completely specified by  background metric. 
Consequently, the only operators that arise 
in the full {\it un-projected} RG flow
in the \lq{}t Hooft limit 
are the single-trace energy-momentum tensor 
and multi-trace operators made of 
the energy-momentum tensor.
Although it is not clear whether such a matrix field theory exists,
we proceed with the assumption that it exists to illustrate
how quantum gravity 
emerges via quantum RG
in a simple setting.
For more general field theories, 
one has to include more operators,
but the generalization is straightforward.

The generating function is written as
$
Z[ g^{(0)}{} ]  =  \int D \Phi ~~ e^{ i S_1[ \Phi; g^{(0)}(x) ] },
$
where $\Phi$ represents underlying microscopic degrees of freedom, and 
$g^{(0)}_{\mu \nu}(x)$ 
with $\mu, \nu = 0,1,..,(D-1)$
is a $D$-dimensional background metric
with signature $(-,+,+,..,+)$.
It is assumed that the regularization scheme respects the $D$-dimensional diffeomorphism invariance.
$S_1$ includes only single-trace operators and
is $O(N^2)$ in the 't Hooft limit.
The energy-momentum tensor,  normalized to be $O(1)$, is given by
$T^{\mu \nu}(x;g^{(0)}] = \frac{1}{N^2 \sqrt{|g^{(0)}|}}\frac{ \delta S_1}{\delta g^{(0)}_{\mu \nu}(x)}$.  
%
Under a coarse graining, the action is modified by a quantum correction
$\delta S^{(1)'}$ which includes 
the Casimir energy,
the single-trace energy-momentum tensor
and double-trace operators constructed from $T^{\mu \nu}$.
The new effective action $S_1[ \Phi;  g^{(0)}(x) ]   +  \delta S^{(1)'}$
 should reproduce the exact same generating function,
\bqa
Z[ g^{(0)} ] & = & \int Dg^{(1)}_{\mu \nu} D\pi^{(1) \mu \nu}
 D \Phi ~~  
e^{ i N^2 \int d^Dx ~ \pi^{(1) \mu \nu} 
( g^{(1)}_{ \mu \nu} - g^{(0)}_{\mu \nu} ) 
}
\nn && \times
e^{ i \delta S^{(1)'}[ i/ N^2 \delta/\delta g^{(1)}_{\mu \nu}, g^{(0)}
 ]} 
e^{i S_1[ \Phi; g^{(1)} ]  },
\label{Z22}
\eqa
where the quantum correction is expressed in terms of functional derivative
with respect to auxiliary fields as in Eq. (\ref{Z01}),
\bqa
&& \delta S^{(1)'}
= 
 dz \int d^Dx ~~ n^{(1)z}(x) \Bigg\{
N^2 \sqrt{|g^{(0)}| } \Bigl( - C_0 
\nn &&
 + C_1 ~ ^{D}{\cal R}(x;g^{(0)}] \Bigr)    
 +i A_{\mu \nu} (x;g^{(0)}] 
 \frac{ \delta  }{\delta g^{(1)}_{\mu \nu}(x) }  
\nn && 
- \frac{ B_{\mu \nu; \rho \sigma }(x;g^{(0)}] }{2N^2} 
\frac{ \delta  }{\delta g^{(1)}_{\mu \nu}(x) }
 \frac{ \delta  }{\delta g^{(1)}_{\rho \sigma}(x) }  
 + ...
\Bigg\}.
\label{dS1}
\eqa
Here we adopt a local RG scheme\cite{OSBORN,SLEE112} 
where the length scale is increased in a spacetime dependent way :
$n^{(1)z}(x)$ is a local speed of coarse graining.
$C_0, ~C_1~ ^{D}{\cal R}$ are the
first two leading order terms of the 
Casimir energy in the derivative expansion, 
where $^{D}{\cal R}$ 
is the $D$-dimensional Ricci scalar\cite{SAKHAROV,VISSER}.
$A_{\mu \nu}$ represents 
the warping factor of the $D$-dimensional spacetime.
$B_{\mu \nu,\rho \sigma}$
represents the source of 
the double-trace operator, 
$T^{\mu \nu} T^{\rho \sigma}$
that is generated under coarse graining.
From dimensional ground, 
we expect $C_0 \sim a^{-D}$, 
$C_1 \sim a^{-D+2}$,
$A_{\mu \nu} \sim 1$ and
$B_{\mu \nu; \rho \sigma } \sim a^{D}$,
where $a$ is a short-distance cut-off scale,
the only scale in the theory.
The ellipsis represents
higher derivative terms
in the Casimir energy
and the contribution of the operators that involve at least one derivative,
such as
$( \nabla_{\alpha_1}^{(1)} \nabla_{\alpha_2}^{(1)} ... ) \frac{ \delta  }{\delta g^{(1)}_{\mu \nu}(x) }$,
$ \frac{ \delta  }{\delta g^{(1)}_{\mu \nu}(x) } 
( {\overleftarrow \nabla_{\alpha_1}^{(1)}} {\overrightarrow  \nabla_{\alpha_2}^{(1)} } ... )
 \frac{ \delta  }{\delta g^{(1)}_{\rho \sigma}(x)}$,
 where $\nabla^{(1)}$ is the covariant derivative with respect to the metric $g^{(1)}_{\mu \nu}$.
The higher derivative terms are suppressed by additional powers of $a$. 

In the local RG, $n^{(1)z}$ is a gauge freedom which controls the local speed of coarse graining.
One can introduce another gauge freedom by using the fact that $Z[ g^{(1)} ]$ is invariant under the $D$-dimensional diffeomorphism,
$Z[ g^{(1)}_{ \mu \nu} ] = Z\left[ g^{(1)}_{ \mu \nu} + dz( \nabla_\mu^{(1)} n^{(1)}_{\nu} + \nabla_\nu^{(1)} n^{(1)}_{\mu} )  \right]$,
where $dz ~ n^{(1)\mu}$ is an infinitesimal shift of the $D$-dimensional coordinates for the low-energy field with respect to the coordinates of the high-energy field.
This leads to
\bqa
&& Z[ g^{(0)} ]  =  \int Dg^{(1)}_{\mu \nu} D\pi^{(1) \mu \nu}
D \Phi ~~  
e^{ i N^2 \int d^Dx ~  \pi^{(1) \mu \nu} 
( g^{(1)}_{ \mu \nu} - g^{(0)}_{\mu \nu} )} \nn
&&
e^{ i \delta S^{(1)'}[i/ N^2 \delta/\delta g^{(1)}_{\mu \nu}; g^{(0)}]} 
e^{ i \delta S^{(1)''}[ i/ N^2 \delta/\delta g^{(1)}_{\mu \nu}]} 
e^{iS_1[ \Phi; g^{(1)}]  }, 
\label{Z221}
\eqa
where
$ \delta S^{(1)''} =  
-i dz~  \int d^Dx  
(  \nabla_\mu^{(1)} n^{(1)}_{\nu} + \nabla_\nu^{(1)} n^{(1)}_{\mu} ) 
 \frac{ \delta }{\delta g^{(1)}_{ \mu \nu}(x)}.$
Integrating $g^{(1)}_{\mu \nu} $ by part,
$\delta S^{(1)'} +  \delta S^{(1)''}$
becomes a quadratic polynomial of $\pi^{(1) \mu \nu}$.
Repeating these steps, the generating function is written as
\bqa
Z[ g^{(0)} ] & = & \int
\prod_{l=1}^L 
\left[ 
Dg^{(l)}_{\mu \nu}(x) 
D\pi^{(l) \mu \nu}(x) 
\right]
 ~~ 
e^{ iS_B } Z[ g^{(L)}(x)], \nn
\label{Z24}
\eqa
where the bulk action, to the linear order of $dz$, becomes
\bqa
&& S_B  = 
N^2 dz \sum_{l=1}^L \int d^Dx ~  
\Bigg\{ 
 \pi^{(l) \mu \nu} \frac{ g^{(l)}_{\mu \nu} - g^{(l-1)}_{\mu \nu}}{dz} 
\nn &&
+ 2  n^{(l)\mu}(x) \nabla^{(l)\nu} \pi^{(l)}_{\mu \nu}  
 +  n^{(l)z} (x,z) \Bigl[
\sqrt{ | g^{(l-1)}  |  } \Bigl( -C_0  \nn 
&& + C_1~ ^{D}{\cal R}(x, g^{(l-1)} ] \Bigr)  
 + A_{\mu \nu}(x;g^{(l-1)}] \pi^{(l) \mu \nu} \nn
&&
+ \frac{ B_{\mu \nu; \rho \sigma }(x;g^{(l-1)}] }{2} 
\pi^{(l) \mu \nu }
\pi^{(l) \rho \sigma } + ...
\Bigr]
\Bigg\}. 
\eqa

In the $dz \rightarrow 0$ limit,
the metric and the conjugate field become $(D+1)$-dimensional fields.
In order to make the radial coordinate more symmetric with the 
$D$-dimensional coordinate $x$, 
we introduce a dimensionful radial coordinate $x^D = a e^{l ~ dz}$,
and define $(D+1)$-dimensional coordinate as $X=(x, x^D)$.
The dimensionless lapse and shift functions are defined as 
$N^\mu(X) \equiv n^{(l) \mu}(x)/x^D$
and
$N^D(X) \equiv n^{(l)z}(x) a/x^D$.
Then the bulk theory takes the form of
a constrained Hamiltonian system 
for the metric and its conjugate momentum\cite{ADM},
\bqa
S_B & = &
N^2 \int dX^{D+1} ~  \Bigl\{ 
 \pi^{\mu \nu} ( \partial_D g_{\mu \nu} )
- N^{\mu} {\cal H}_\mu  - N^D {\cal H}
\Bigr\}, \nn
\eqa
where the momentum and Hamiltonian constraints are given by
\bqa
{\cal H}_\mu  &=&  - 2 \nabla^{\nu} \pi_{\mu \nu}, \nn
{\cal H}  &=&   
-\gamma \sqrt{|g|}( -\Lambda_0 +  ^{D}{\cal R}  )
-  \beta_{\mu \nu}  \pi^{\mu \nu}  \nn
&&  - \frac{{\cal G}_{\mu \nu; \rho \sigma }}{2}
\pi^{\mu \nu} \pi^{\rho \sigma} + ... 
\label{HH2}
\eqa
with 
$\Lambda_0 = C_0/C_1$,
$\gamma = C_1/a$,
$\beta_{\mu \nu} = A_{\mu \nu}/a$,
and
${\cal G}_{\mu \nu; \rho \sigma} = B _{\mu \nu; \rho \sigma} /a$.
The most general forms of the warping factor
and the super-metric which are consistent with the $D$-dimensional
diffeomorphism invariance are
\bqa
\beta_{\mu \nu} & = & \beta g_{\mu \nu}, \nn
{\cal G}_{\mu \nu; \rho \sigma } &=& 
 \frac{\alpha}{\sqrt{|g|} } ( g_{\mu \rho} g_{\nu \sigma}
- \lambda  g_{\mu \nu} g_{\rho \sigma} )
\eqa
to the leading order in $a$.
For unitary theories, $\alpha,  \beta, \gamma, \Lambda_0$ are real,
and their values depend on the matter content\cite{VISSER}.
Here we focus on the case with $\alpha, \gamma > 0$. 
In the bulk action, these parameters always appear in combination with $N^z$.
Therefore one can choose $N^z$ 
to normalize one of the parameters.
This freedom stems from the fact that 
only relative speeds of RG 
flows for different operators matter. 
One often chooses $N^z$ 
to fix the warping factor as 
$\beta= 2/a$\cite{OSBORN}.
Here we make an alternative choice such that 
$\frac{\alpha}{2} = \frac{1}{\gamma} \equiv 2 \kappa^2 \sim a^{D-1} $.
As expected, the warping term, $\beta_{\mu \nu} \pi^{\mu \nu}$ breaks the SR symmetry. 
However, $\beta_{\mu \nu}$ can be written as a gradient flow
$\beta_{\mu \nu}  =   {\cal G}_{\mu \nu, \rho \sigma} \frac{\delta c}{\delta g_{\rho \sigma}(x)}$ 
with $c=  -\frac{\beta}{2 \kappa^2 (D\lambda - 1)} \int d^Dx \sqrt{|g|}$.
The SR-symmetry breaking term is traded with 
a cosmological constant $\frac{1}{2} \frac{\delta c}{\delta g_{\mu \nu}}  {\cal G}_{\mu \nu; \rho \sigma}  \frac{\delta c}{\delta g_{\rho \sigma}} $
in the bulk and 
the surface tension term $N^2 \left( c[g(x,x^D=a)] - c[g(x,x^D=\infty)] \right)$ at the boundaries.

In order to fix $\lambda$,
we note that ${\cal H}=0$ and ${\cal H}^\mu=0$ on shell 
because the generating function is independent of the choice 
of $\{ N^\mu(X),N^D(X) \}$ :
choice of different RG prescriptions is a pure gauge freedom
which does not affect the generating function.
They have to be of first class
because the constraint equations ${\cal H}=0$, ${\cal H}^\mu=0$ 
are satisfied on shell at any $X$ in the bulk
for any choice of $\{ N^\mu(X),N^D(X) \}$\cite{SLEE112}.
The fact that they should form first-class constraints, 
which holds at each order in $a$,
forces $\lambda = \frac{1}{D-1}$.
In $D=3$, this was shown in Ref. \cite{HENN}.
It is straightforward to extend the result to general dimensions.

Once the canonical momentum is integrated out,
one obtains the $(D+1)$-dimensional Einstein gravity
upto the two derivative terms,
\bqa
S_B & = & \frac{N^2}{2 \kappa^2} \int d^{D+1}X ~ \sqrt{|G|}
\Bigl(
-\Lambda  + ~^{(D+1)}{\cal R} + ..
\Bigr).
\label{EG}
\eqa
Here 
$G_{M N}$ with $M,N=0,1,2,...,D$ is the 
$(D+1)$-dimensional metric constructed from
$g_{\mu \nu}$, $N^\mu$ and $N^D$
with signature $(-,+,+,..,+)$.
The signature of the radial direction is determined by
the sign of $\alpha$\footnote{
If $\alpha$ was negative, the radial direction
would be time-like.}.
$\Lambda = \Lambda_0 - \frac{D(D-1) \beta^2}{4}$
is the cosmological constant.
$\Lambda_0$ is determined by the 
vacuum energy density of the field theory per unit
RG length scale, that is, the contribution
from the modes integrated between the length scale
$a$ and $a e^{dz}$. 
If $\Lambda_0=0$, 
which holds for supersymmetric theories,
the cosmological constant is negative
as it is solely determined by
the warping factor.
In this case, one naturally obtains AdS in the bulk.
The Newton constant is $G_N \sim \kappa^2/N^2$, 
where $\kappa^2$ is given by the ratio between 
the beta function of the double-trace energy-momentum tensor
and the coefficient of $^D {\cal R}$ in the Casimir energy. 
The ellipsis in Eq. (\ref{EG}) denotes terms that have more than two derivatives.
Generically, $\Lambda$, $\kappa^2$
and the scale for the higher derivative terms 
will be determined by the UV cut-off scale
and dynamical properties of the boundary quantum field theory.
One obtains a weakly curved spacetime in the bulk
if the cosmological constant is smaller than 
the scales associated with the higher derivative terms, 
as is the case for the ${\cal N}=4$ super Yang-Mills theory
in the strong coupling limit.
It will be of great interest to understand the precise condition
under which the weakly curved spacetime emerges in the bulk.


Finally, we comment on holographic duals for (ungauged) vector models\cite{KP}.
Once multi-trace operators are traded with dynamical sources in the first step of RG as in Eq. (\ref{Z02}), 
the resulting single-trace action remains quadratic in the following steps\cite{Douglas:2010rc}.
This generates a bulk action which is linear in the conjugate momenta\cite{DOLAN} except at the UV boundary. 
This corresponds to the zero coupling limit (say $\kappa =0$) in the bulk.

We thank 
Hong Liu, 
Yu Nakayama,
Joao Penedones, Joe Polchinski,
Renate Loll, Sang-Jin Sin 
and Tadashi Takayanagi
for helpful comments.
This research was supported by ERA, NSERC and the Templeton Foundation.
Research at the Perimeter Institute is supported 
in part by the Government of Canada 
and by the Province of Ontario.

\end{document}